\documentclass[aps,prb,twocolumn,10pt,superscriptaddress,nofootinbib]{revtex4-2}

\usepackage{amsmath, amssymb}
\usepackage{mathtools}
\usepackage{physics}
\usepackage{graphicx}
\usepackage{hyperref}
\usepackage{slashed}
\usepackage{orcidlink}
\usepackage[dvipsnames,x11names]{xcolor}
\usepackage{bbm}
\usepackage{mathtools}
\usepackage{newtxtext}
\usepackage[varvw]{newtxmath}
\usepackage{placeins} 
\newcommand{\Lagr}{\mathcal{L}}
\usepackage{parskip}

\hypersetup{
    colorlinks,
    linkcolor={blue},
    citecolor={blue},
    urlcolor={blue}
}

\usepackage{color}
\usepackage[normalem]{ulem}
\usepackage[export]{adjustbox}
\usepackage{bbold}
\allowdisplaybreaks[1]

\begin{document}

\title{Spontaneous symmetry breaking of $\mathrm{SO}(2N)$ in Gross--Neveu theory from $2+\epsilon$ expansion}

\author{Bilal Hawashin\,\orcidlink{0009-0005-6335-0483}}
\email{hawashin@tp3.rub.de}
\affiliation{Theoretische Physik III, Ruhr-Universit\"at Bochum, D-44801 Bochum, Germany}

\author{Max Uetrecht\,\orcidlink{0000-0001-8685-2543}}
\email{max.uetrecht@tu-dortmund.de}
\affiliation{Fakult\"at Physik, Technische Universit\"at Dortmund, D-44221 Dortmund, Germany}

\begin{abstract}
It was recently established that the paradigmatic Gross--Neveu model with $N$ copies of two-dimensional Dirac fermions features an $\mathrm{SO}(2N)$ symmetry if certain interactions are suppressed. This becomes evident when the theory is rewritten in terms of $2N$ copies of two-dimensional Majorana fermions. 
Mean-field theory for the $\mathrm{SO}(2N)$ model predicts, besides the chiral Ising transition at $g_{c1}$, a second critical point $g_{c2}$ where $\mathrm{SO}(2N)$ is broken down to $\mathrm{SO}(N)\times\mathrm{SO}(N)$. A subsequent Wilsonian renormalization group analysis directly in $d=3$ supports its existence in a generalized theory, where $N_f$ copies of the $4N$-component Majorana fermions are introduced. This allows to track the evolution of a (i) quantum anomalous Hall Gross--Neveu--Ising, (ii) symmetric-tensor, and (iii) adjoint-nematic fixed point separately. However, it turns out that (ii) and (iii) lose their criticality when approaching $N_f = 1$, suggesting that the transition is first order.
In this work, we approach the problem from the lower-critical dimension of two. We construct a Fierz-complete renormalizable Lagrangian, compute the leading order $\beta$ functions, fermion anomalous dimension, as well as the order parameter anomalous dimensions, and resolve the three universality classes corresponding to (i)--(iii). 
Before becoming equal to the Gaussian fixed point at $N_f = 1$, (ii) remains critical for all values of $N_f > N_{f,c}^{\mathrm{ST}}(N) \approx 0.56 + 1.48 N +\mathcal{O}(\epsilon)$, which compares well with the estimate of previous studies. 
We further find that (iii) becomes equal to (i) when approaching $N_f = 1$. An instability is, however, only present in the susceptibility corresponding to the Gross--Neveu--Ising order parameter.
\end{abstract}

\maketitle

\section{Introduction} 
Dirac materials~\cite{Vafek2014,Wehling2014} denotes a class of materials whose low-energy excitations are given by massless Dirac fermions. The primary example of a Dirac material is graphene, where the energy dispersion becomes linear close to the two inequivalent Dirac points~\cite{Herbut:2006,Herbut:2009}. Low-energy excitations of such materials can therefore be effectively described by a relativistic quantum field theory, allowing for a systematic study of interaction-driven phase transitions in such systems. Interacting field theories that arise in this context are known as Gross--Neveu models~\cite{Gross:1974,Bondi:1989gba,Gracey:1990sx,Herbut:2006,PhysRevB.79.085116,Gehring:2015,Gracey:2021yyl,Gracey:2025aoj}, and they have been shown to feature various interacting fixed-points between two and four spacetime dimensions. By using a Hubbard--Stratonovich transformation, four-fermion interactions can be exactly rewritten in terms of a Yukawa interaction of the fermions with a bosonic field that naturally plays the role of an order parameter field, which transforms in an irreducible representation (irrep) of the underlying global symmetry group.
A finite vacuum expectation value (vev) of the order parameter field in the infrared therefore signals spontaneous symmetry breaking of the corresponding global symmetry, and dynamically gaps out the Dirac fermions. The transition of a Dirac (semi)metal into an ordered insulating phase is known as relativistic Mott transition~\cite{Herbut:2006,Herbut2:2009}. Due to the insufficient interaction strength in graphene, such a transition can not be observed in the paradigmatic Dirac material. 

Recently, a tunable platform for the engineering of electronic properties has emerged, known as moiré materials, see, e.g., Refs.~\cite{he2021moire,mak2022semiconductor,shi2021exotic} for an overview. The primary example of a moiré material is twisted bilayer graphene~(TBG), in which two single layers of graphene are stacked on top of each other with a relative twist between them. When tuning the twist angle to a value of around $1.1^\circ$~\cite{bistritzer2011moire}, the energy bands become nearly flat and the bandwidth is therefore vanishingly small. Evidently, interaction effects become enhanced and various correlated phases emerge~\cite{bistritzer2011moire,cao2018correlated,lu2019superconductors}. In fact, TBG hosts Dirac excitations right at charge neutrality~\cite{bistritzer2011moire,PhysRevB.108.235120}, therefore providing an experimentally accessible platform for the realization of the relativistic Mott transition. Recently, the relativistic Mott transition upon tuning the twist angle has been experimentally observed for the first time in twisted bilayer WSe${}_2$~\cite{ma2024relativistic}, and has been argued to belong to the Gross--Neveu--Heisenberg universality class~\cite{ma2024relativistic,biedermann2025dirac,hawashin2025relativistic}.

A recent series of works~\cite{Herbut:2023une,Han:2024ird,PhysRevB.110.125131,Han:2025kjt} has established that the Gross--Neveu model with global $\mathrm{SU}(N) \times \mathrm{U}(1)$ symmetry has a symmetry-enhancement to $\mathrm{SO}(2N)$ if certain instabilities are suppressed, which becomes manifest if the theory is rewritten in terms of Majorana fermions. The case of hopping electrons on the honeycomb lattice correspond to $N = 4$, and it has been argued that the $\mathrm{SO}(8)$ symmetry unifies all order parameters within the symmetric-tensor representation, except the $\mathrm{SO}(8)$ singlet whose nonvanishing vev corresponds to the quantum anomalous Hall state~\cite{Herbut:2023une}. The case of $N=8$ corresponds to twisted bilayer graphene, which has double the amount of degrees of freedom due to the additional mini-layer structure. The higher $\mathrm{SO}(2N)$ symmetry enlarges the theory space and allows for new critical points that possibly separate a $\mathrm{SO}(2N)$ symmetric phase from an ordered phase, where $\mathrm{SO}(2N)$ is spontaneously broken. In fact, the existence of such critical points has been shown at the mean-field level~\cite{Herbut:2023une}: (i) Gross--Neveu--Ising critical point, corresponding to the quantum anomalous Hall~(QAH) state with a nonvanishing vev of the $\mathrm{SO}(2N)$ singlet, (ii) a symmetric tensor critical point, separating the $\mathrm{SO}(2N)$ symmetric from a $\mathrm{SO}(2N) \to \mathrm{SO}(N) \times \mathrm{SO}(N)$ broken phase, and (iii) an adjoint-nematic critical point, where $\mathrm{SO}(2N)$ and Lorentz symmetry are spontaneously broken. 

Subsequently, the corresponding fixed points have been identified within a Wilsonian renormalization group (RG) analysis~\cite{Han:2024ird} of a generalized Gross--Neveu model in the large-$N_f$ limit, where $N_f$ copies of the $4N$-component Majorana fermions are introduced. However, it turns out that the fixed points (ii) and (iii) imply a diverging susceptibility for its corresponding order parameter only for $N_f(N) \approx 0.35N$~\cite{Herbut:2023une}.
As the original Gross--Neveu model is recovered for $N_f = 1$, this can be interpreted as the transition into the $\mathrm{SO}(2N)$ broken phase being of first order for all $N > 2$.
A complementary study of the corresponding Gross--Neveu--Yukawa theory for symmetric tensor criticality has also been carried out with $(4-\epsilon)$--expansion up to two-loop and partly up to three-loop order~\cite{PhysRevB.110.125131,Han:2025kjt}. At leading order, the Gross--Neveu--Yukawa theory shows the existence of two critical flavors numbers $N_{f,c1} < N_{f,c2}\,$: for $N_f > N_{f,c2}$, a critical point exists that separates the $\mathrm{SO}(2N)$ symmetric phase from an ordered phase with spontaneous breaking of $\mathrm{SO}(2N) \to \mathrm{SO}(N) \times \mathrm{SO}(N)$. For $N_{f,c1} < N_f < N_{f,c2}$, the critical point lies in a region of parameter space where the effective potential is unbounded from below, and finally it collides with another fixed point for $N_f = N_{f,c1}$, rendering the symmetric tensor fixed point complex for $N < N_{f,c1}$. At leading order, $N_{f,c1} \approx N$ and $N_{f,c2} \approx 2N$~\cite{Han:2024swe}. Higher-loop corrections corroborate these findings, however, the $4-\epsilon$ expansion seems to be divergent and requires additional resummation schemes~\cite{Han:2025kjt}. The absence of a fixed point for $N_f < N_{f,c1}$ implies the transition to be of first order, compatible with the results from Wilsonian RG. However, the mechanism leading to the absence of symmetric-tensor criticality in $d=2+1$ is completely different to the, rather unconventional, mechanism observed with Wilsonian RG. A crucial question is how the seemingly different mechanisms can be understood consistently, and whether they are artifacts of the respective approaches.

In order to understand the fixed-point structure in $d = 2+1$ dimensions better, we tackle the problem from the lower critical dimension and perform a Fierz-complete renormalization-group study close to the lower critical dimension of two. In Sec.~\ref{sec:model}, we first show that the global symmetry of the Gross--Neveu theory is enhanced to $\mathrm{SO}(2N)$ if certain interactions are tuned to zero by reformulating the theory in $2 + \epsilon$ spacetime dimensions in terms of Majorana fermions. We then construct a basis of symmetry-allowed four-fermion interactions, and, in Sec.~\ref{sec:rg}, calculate the one-loop $\beta$ functions as well as the fermion anomalous dimension at two-loop order. In Sec.~\ref{sec:critExp}, we identify fixed points corresponding to QAH Gross--Neveu--Ising~(i), symmetric-tensor~(ii), and adjoint-nematic~(iii) criticality, provide estimates of various critical exponents, and estimate a critical flavor number $N_{f,c}^{\mathrm{ST}}$, below which the symmetric-tensor fixed point does not imply a corresponding diverging susceptibility. We further interpolate our result for $N_{f,c}^{\mathrm{ST}}$ from the lower critical dimension with the results from the upper critical dimension available in the literature. Finally, we conclude in Sec.~\ref{sec:conclusion}.

\section{Model}\label{sec:model}

We consider the Gross--Neveu model~\cite{Gross:1974,PhysRevD.10.3235,Bondi:1989gba,BONDI1989345,Herbut:2006,Herbut:2023une,Han:2024ird,Gracey:2025aoj} in $d=2+\epsilon$ Euclidean spacetime dimensions defined by the microscopic action ${S = \int d^dx\, \mathcal{L}}$ with
\begin{align} \label{eq:lagrGN}
    \mathcal{L} = \bar{\psi} (\mathbb{1}_N \otimes \gamma_\mu) \partial_\mu \psi - \frac{g_1}{2} (\bar{\psi} (\mathbb{1}_N \otimes \mathbb{1}_2) \psi)^2\,.
\end{align}
In the above, $\mu = 0, ..., d-1$ denote the spacetime index, and the matrices $\gamma_\mu$ form a $d_\gamma$-dimensional representation of the Euclidean Clifford algebra, i.e., $\{\gamma_\mu, \gamma_\nu\} = 2 \delta_{\mu \nu} \mathbb{1}_{d_\gamma}$. The Dirac fermion $\psi$ and its conjugate $\bar{\psi} = \psi^\dagger (\mathbb{1}_N \otimes \gamma_0)$ are $N\cdot d_\gamma$-component spinors. In $d=1+1$ and $d=2+1$ dimensions, irreducible representations of the Clifford algebra are two-dimensional, i.e., $d_\gamma=2$, and, in the following, we choose a Hermitian and irreducible representation. In $1+1$ dimensions, this results in an additional Lorentz scalar, given by $\gamma_5:=i\gamma_0 \gamma_1$. The Gross--Neveu model defined by Eq.~\eqref{eq:lagrGN} admits an interacting fixed-point in $2 < d < 4$ spacetime dimensions, that has been studied with various methods, such as the $\epsilon$-expansion~\cite{Gross:1974,Bondi:1989gba,Gehring:2015,Zerf:2017,Ihrig:2018}, large-$N$~\cite{Herbut:2006,Gracey:2016mio,Gracey:2025aoj}, and the functional renormalization group~\cite{Janssen:2014,Scherer:2012fjq,Tolosa-Simeon:2025fot}.
In $d=2+1$, the Gross--Neveu and complementary Gross--Neveu--Yukawa theory are expected to flow to the same fixed point, realizing a common universality class that can be probed, e.g., using the $\epsilon$-expansion from both the lower and upper critical dimension~\cite{Focht:1995ie,Gracey:2025aoj}.

The model defined by Eq.~\eqref{eq:lagrGN} is known to have various global symmetries:
\paragraph{Rotational invariance:} In two Euclidean spacetime dimensions, spacetime coordinates transform as $x^\mu \mapsto x'^\mu = (\Lambda^{-1})^{\mu}_{\;\nu} x^\nu$ under spacetime rotations $\Lambda \in \mathrm{O}(2)$. In the Dirac representation, spinors transform as
\begin{align}
    \psi(x) \mapsto (\mathbb{1}_N \otimes \mathcal D(\Lambda)) \psi(x'),
\end{align}
where
$\mathcal D(\Lambda) = e^{-\frac{i}{4} {\omega} \sigma_{01}}$ and $\sigma_{01} = \frac{i}{2} [\gamma_0, \gamma_1]$. 
\paragraph{$\mathrm{U}(1)$ charge symmetry:} For $\alpha \in \mathbb{R}$,
\begin{equation}
    \psi \to e^{i \alpha} \psi\,, \quad \bar{\psi} \to e^{-i \alpha}\bar{\psi}\,,
\end{equation}
implying that any term in the Lagrangian has to contain an equal number of Dirac fermions and their conjugates. On the lattice, this ensures particle number conservation and charge neutrality.
\paragraph{$\mathrm{SU}(N)$ symmetry:} For $U \in \mathrm{SU}(N)$, 
\begin{equation}
    \psi \to (U \otimes 1_2) \psi, \quad \bar{\psi} \to \bar{\psi} (U^\dagger \otimes 1_2).
\end{equation}
All together, this forms the global $\mathrm{SU}(N) \times \mathrm{U}(1)$ symmetry of the Gross--Neveu theory space.

The Lagrangian in Eq.~\eqref{eq:lagrGN} furthermore has various discrete symmetries, such as 
\paragraph{Parity symmetry:} Under parity symmetry, spacetime coordinates transform as $(x_\mu) = (x_0,x_1) \to (x'_\mu) = (x_0,-x_1)$, while the Dirac spinor transforms as
\begin{align}
    \psi(x) \to (\mathbb{1}_N \otimes \gamma_0) \psi(x').
\end{align}
\paragraph{$\mathbb{Z}_2$ chiral symmetry:} Due to the additional Lorentz scalar $\gamma_5$ that anticommutes with all $\gamma_\mu$, the Lagrangian is invariant under the discrete ``chiral'' transformation
\begin{align}
    \psi \to (\mathbb{1}_N \otimes \gamma_5) \psi, \quad \bar{\psi} \to -\bar{\psi} (\mathbb{1}_N \otimes \gamma_5).
\end{align}
We remark that the $\mathbb{Z}_2$ chiral symmetry is anomalous in dimensionally regulated theories due to the well-known ``$\gamma_5$ problem''. Here, we avoid an explicit treatment of $\gamma_5$ by choosing a Dirac basis without it, cf., App.~\ref{app:martin}.

Any four-fermion theory that features the above symmetries belong to the so-called Gross--Neveu theory space~\cite{Gehring:2015,Herbut:2023une}, and the theory defined by Eq.~\eqref{eq:lagrGN} is only one representative of this theory space. In fact, the symmetries listed above admit six interaction terms
\begin{align}\label{eq:LintSUN}
    \mathcal{L}_\text{int} &= - \frac{g_1}{2} (\bar{\psi} (\mathbb{1}_N \otimes \mathbb{1}_2) \psi)^2 - \frac{g_2}{2} (\bar{\psi} (\mathbb{1}_N \otimes \gamma_\mu) \psi)^2 \notag \\ 
    &\quad -\frac{g_3}{2} (\bar{\psi} (\mathbb{1}_N \otimes \gamma_5) \psi)^2 - \frac{g_4}{2} (\bar{\psi} (T^a \otimes \mathbb{1}_2) \psi)^2 \notag \\
    &\quad -\frac{g_5}{2} (\bar{\psi} (T^a \otimes \gamma_\mu) \psi)^2 - \frac{g_6}{2} (\bar{\psi} (T^a \otimes \gamma_5) \psi)^2,
\end{align}
where $T^a$ with $a=1,...,N^2-1$ are the generators of $\mathrm{SU}(N)$. The basis above reduces to four interactions in $d=2+1$, where $g_3 = g_2$ and $g_5 = g_6$ in order to maintain Lorentz invariance~\cite{Herbut:2023une}. The six interactions terms above are further related by Fierz identities, see, e.g., Refs.~\cite{Herbut:2009,Herbut:2023une} and Sec.~\ref{sec:Fierz}.
As noted in Refs.~\cite{Herbut:2023une,Han:2024ird,Han:2025kjt,PhysRevB.110.125131}, if the interactions are tuned such that only the first four-fermion term is nonvanishing, i.e., if we consider the canonical Gross--Neveu model defined by Eq.~\eqref{eq:lagrGN}, the theory actually has a larger $\mathrm{SO}(2N) \supset \mathrm{SU}(N) \times \mathrm{U}(1)$ symmetry. This symmetry is hidden but of course present for general representations of the Clifford algebra. 

In order to make this $\mathrm{SO}(2N)$ symmetry manifest, we consider the class of representations for which $\gamma_0^* = -\gamma_0$, and $\gamma_i^* = \gamma_i$ for $i=1,5$, which is real as well as chiral, and decompose the complex $2N$-component Dirac spinor into two $2N$-component Majorana fermions $\chi_{1,2}$ via
\begin{align}
    \psi = (\chi_1 + i \chi_2)/\sqrt{2}, \quad \bar{\psi} = (\bar{\chi}_1 - i \bar{\chi}_2)/\sqrt{2},
\end{align}
with $\bar{\chi}_i = \chi_i^T (\mathbb{1}_N \otimes \gamma_0)$. The Majorana fields $\chi_i$ are then ``real'' Grassmann fields, and the Lagrangian~\eqref{eq:LintSUN} becomes
\begin{align}\label{eq:majlagr}
    \mathcal{L} = \frac{1}{2}\bar{\chi} (\mathbb{1}_{2N} \otimes \gamma_\mu) \partial_\mu \chi - \frac{g_1}{8} (\bar{\chi} (\mathbb{1}_{2N} \otimes \mathbb{1}_2) \chi)^2\,,
\end{align}
with the $4N$-component spinors $\chi := (\chi_1^T,\chi_2^T)^T$ and ${\bar{\chi} := (\bar{\chi}_1,\bar{\chi}_2)}$.
It is crucial to note that for the chosen class of representations of $\gamma$ matrices, $\bar{\chi}_1 \chi_2 = \bar{\chi}_2 \chi_1$ (recall that $\gamma_0^* = -\gamma_0$), and mixed terms of the form $\bar{\chi}_i (\mathbb{1}_{N} \otimes \gamma_\mu) \partial_\mu \chi_j$ with $i \neq j$ cancel after partial integration. We stress again that the theory described by Eq.~\eqref{eq:majlagr} is the same as the theory described by Eq.~\eqref{eq:lagrGN}. It is now manifest that the Lagrangian in Eq.~\eqref{eq:majlagr} is invariant under the transformation
\begin{equation}
    \chi \to (O \otimes \mathbb{1}_2) \chi, \quad \bar{\chi} \to \bar{\chi} (O^T \otimes \mathbb{1}_2)
\end{equation}
with $O \in \mathrm{SO}(2N)$, i.e., $\chi$ transforms as a vector under $\mathrm{SO}(2N)$.

In the following, we will generalize the model defined by Eq.~\eqref{eq:majlagr} by introducing $N_f$ flavors of Majorana fermions $\chi$, similar to Refs.~\cite{Han:2024ird,Han:2025kjt}, i.e., we consider
\begin{align}\label{eq:majLagr}
    \mathcal{L} = \frac{1}{2} \bar{\chi}_k (\mathbb{1}_{2N} \otimes \gamma_\mu) \partial_\mu \chi_k - \frac{g_1}{8} (\bar{\chi}_k (\mathbb{1}_{2N} \otimes \mathbb{1}_2) \chi_k)^2,
\end{align}
where $k=1,...,N_f$. This promotes the $\mathrm{SO}(2N)$ symmetry to $\mathrm{SO}(2N) \times \mathrm{SO}(N_f)$, and will allow us to investigate a large-$N_f$ limit on the level of the $\beta$ functions that majorly simplifies the fixed-point analysis.
Furthermore, the enlarged $\mathrm{SO}(2N)$ symmetry gives rise to new four-fermion interaction terms and fixed points, which can be tracked independently for $N_f > 1$.
Note that if we had introduced the flavor structure already in Eq.~\eqref{eq:lagrGN}, the global symmetry would have simply become $\mathrm{SO}(2N N_f)$.

\section{Renormalization group} \label{sec:rg}

In $d$ spacetime dimensions, four-fermion couplings $g_i$ have mass dimension $[g_i] = d - 2$. Slightly above two spacetime dimensions, the coupling turns irrelevant, and the theory becomes perturbatively renormalizable. This identifies $d=1+1$ as the lower critical dimension, see also Refs.~\cite{Bondi:1989gba,BONDI1989345,Gehring:2015,Ladovrechis:2023}. In $d=2+\epsilon$, the $\beta$ function of the Gross--Neveu coupling $g_1$ in Eq.~\eqref{eq:majlagr} schematically reads $\beta_{g_1} = \epsilon g_1 - c g_1^2$ with $c > 0$, and yields an interacting fixed point $g_1^* = \epsilon/c$ that emerges from the Gaussian one at $\mathcal{O}(\epsilon)$. In the presence of such a fixed point, $g_1$ can be tuned above the finite value of $g_1^*$, which then leads to a runaway flow toward positive infinity~\cite{Kaveh:2004qa,Hawashin:2025}.
Such a runaway flow generally hints toward spontaneous symmetry breaking, where the divergence of the coupling signals the dynamical generation of a nonvanishing order parameter.
However, in strictly two spacetime dimensions, the spontaneous breaking of continuous global symmetries is excluded~\cite{Coleman:1973ci},
and a runaway in a direction associated with continuous symmetry breaking cannot be taken as evidence for true long-range order.
Instead, the renormalization-group (RG) flow enters a strong coupling regime, where bosonic order-parameter fluctuations---absent in our present analysis---must be taken into account~\cite{Witten:1978qu,Rosenstein:1990nm,Barducci:1994cb}.

Given that the RG flow generically generates additional symmetry-allowed interaction terms, we proceed with a Fierz-complete study of purely fermionic theories.

\subsection{Symmetry-allowed operators}

We now identify all four-fermion terms allowed by symmetries of the theory defined in Eq.~\eqref{eq:majLagr}. Bilinears must be of the form $(\bar{\chi}_k \mathcal{O} \chi_k)$ due to charge symmetry, where $\mathcal{O} = \mathcal{M} \otimes \mathcal{N}$ is a $4N \times 4N$ dimensional operator, with $\mathcal{N}$ acting in the space of $2N \times 2N$ matrices and $\mathcal{M}$ acting in the space of $2 \times 2$ matrices. A basis $\mathcal{B}$ in the space of $4N \times 4N$ matrices is given by
\begin{align}\label{eq:basis2D}
    \mathcal{B} = \mathcal{B}_1 \otimes \mathcal{B}_2 = \{\mathbb{1}_{2N}, \mathbb{A}_a, \mathbb{S}_b\} \otimes \{\mathbb{1}_2,\gamma_{\mu}, \gamma_5 \},
\end{align}
where $a=1,...,N(2N-1)$ and $b=1,...,(N+1)(2N-1)$. 
The matrices $\mathbb{A}_a$ are the antisymmetric generators of $\mathrm{SO}(2N)$ with $\mathrm{Tr}(\mathbb{A}_a \mathbb{A}_{a'}) = 2N \, \delta_{aa'}$ and transform in the adjoint representation of $\mathrm{SO}(2N)$, while the matrices $\mathbb{S}_b$ are symmetric and traceless with $\mathrm{Tr}(\mathbb{S}_b \mathbb{S}_{b'}) = 2N \, \delta_{bb'}$ and transform as symmetric tensors under $\mathrm{SO}(2N)$.

All possible four-fermion terms can be constructed from bilinears. They can be taken as flavor singlets, i.e., $(\bar{\chi}_k \mathcal{O} \chi_k)(\bar{\chi}_l \mathcal{Q} \chi_l)$, or flavor nonsinglets, i.e., $(\bar{\chi}_k \mathcal{O} \chi_l)(\bar{\chi}_l \mathcal{Q} \chi_k)$, where $\mathcal{O}, \mathcal{Q} \in \mathcal{B}$. However, flavor singlets and nonsinglets can be related by the Fierz rearrangement-formula~\cite{PhysRevB.79.085116}. In the case of trivial flavor structure, that is, $N_f = 1$, those Fierz identities yield additional linear relations among four-fermion operators, further reducing the number of linearly independent terms. In the following, we will first consider the case of $N_f > 1$, i.e., all flavor singlets are linearly independent, and discuss the case of trivial flavor structure separately.

Due to the $\mathrm{SO}(2N)$ symmetry, interaction terms must be of the form $(\bar{\chi} (\mathcal{M} \otimes \mathcal{N}_1) \chi)(\bar{\chi} (\mathcal{M} \otimes \mathcal{N}_2) \chi)$, where $\mathcal{M} \in \mathcal{B}_1$ and ${\mathcal{N}_i \in \mathcal{B}_2}$. Lorentz symmetry and parity symmetry further imply that $\mathcal{N}_1 = \mathcal{N}_2$, and hence all four-fermion terms are of the form $(\bar{\chi} (\mathcal{M} \otimes \mathcal{N}) \chi)^2$. Further, note that some terms constructed in this way are vanishing. This is because of the additional identity (recall that in the chosen class of representations, $\gamma_0^T = -\gamma_0$),
\begin{align}\label{eq:MajoranaSelecRule}
    \bar{\chi}_k (\mathcal{M} \otimes \mathcal{N}) \chi_k &= \chi_k^T (\mathcal{M} \otimes \gamma_0 \mathcal{N}) \chi_k \notag\\ 
    &= \chi_k^T (\mathcal{M}^T \otimes \mathcal{N}^T \gamma_0) \chi_k \,.
\end{align}
Hence, a bilinear is nonvanishing only in the following cases: (1) if $[\gamma_0,\mathcal{N}] = 0$, then $(\mathcal{M} \otimes \mathcal{N})^T = \mathcal{M} \otimes \mathcal{N}$, and (2) if $\{\gamma_0,\mathcal{N}\} = 0$, then $(\mathcal{M} \otimes \mathcal{N})^T = -\mathcal{M} \otimes \mathcal{N}$. In total, a Fierz-complete basis of linearly independent four-fermion terms for $N_f > 1$ is given by 
\begin{align}\label{eq:basiseucl}
    \mathcal{L}_\mathrm{int} =& -\frac{\bar{g}_1}{8} (\bar{\chi}_k(\mathbb{1}_{2N} \otimes \mathbbm{1}_2)\chi_k)^2 - \frac{\bar{g}_2}{8} (\bar{\chi}_k (\mathbb{S}_a \otimes \mathbbm{1}_2) \chi_k)^2 \notag\\
    &- \frac{\bar{g}_3}{8}(\bar{\chi}_k (\mathbb{A}_b \otimes \gamma_{\mu}) \chi_k)^2 
    -\frac{\bar{g}_4}{8} (\bar{\chi}_k (\mathbb{A}_b \otimes \gamma_5) \chi_k)^2.
\end{align}
We emphasize that the theory has four linearly independent four-fermion interactions, irrespective of the chosen class of spinor representation. 
We also note that fermion flavor is conserved along any fermion line, implying that the flavor singlets in Eq.~\eqref{eq:basiseucl} cannot induce nonsinglets under RG transformations. This can readily be seen by rewriting, e.g., the first term in Eq.~\eqref{eq:basiseucl}
in terms of a Hubbard--Stratonovich scalar field $\sigma$ that does not carry any flavor indices, i.e., $-\tfrac{g_1}{8} (\bar{\chi}_k (\mathbb{1}_{2N} \otimes \mathbb{1}_2) \chi_k)^2 \rightarrow \tfrac{2}{g_1} \sigma^2 - \bar{\chi}_k (\mathbb{1}_{2N} \otimes \mathbb{1}_2) \chi_k \sigma$.

The Fierz-complete basis given above can be compared to its three-dimensional analog given in Ref.~\cite{Han:2024ird}, where $\bar{g}_4 = \bar{g}_3$ due to Lorentz invariance.

\subsection{Fierz identities for $N_f = 1$}
\label{sec:Fierz}

In the case of trivial flavor structure, i.e., $N_f = 1$, the four-fermion terms in Eq.~\eqref{eq:basiseucl} can be further related through the Fierz rearrangement formula~\cite{Herbut:2009,Herbut:2023une}
\begin{align}
    (\bar{\chi} \mathcal{O} \chi)^2 = -\frac{1}{16 N^2} \mathrm{Tr}(\mathcal{O} X_a \mathcal{O} X_b) (\bar{\chi} X_b \chi)(\bar{\chi} X_a \chi),
\end{align}
where $\mathcal{O}$ is any $4 N \times 4 N$ operator, and $X_a \in \mathcal{B}$ with $\mathrm{Tr}(X_a X_b) = 4N \delta_{ab}$. Defining the column 
\begin{align}
    V = \begin{pmatrix}
        (\bar{\chi}(\mathbb{1}_{2N} \otimes \mathbbm{1}_2)\chi)^2 \\
        (\bar{\chi} (\mathbb{S}_a \otimes \mathbbm{1}_2) \chi)^2 \\
        (\bar{\chi} (\mathbb{A}_b \otimes \gamma_{\mu}) \chi)^2 \\
        (\bar{\chi} (\mathbb{A}_b \otimes \gamma_5) \chi)^2
    \end{pmatrix},
\end{align}
the Fierz rearrangement formula yields the identity $FV = 0$, with the $4\times4$ matrix 
\begin{align}
    F = \begin{pmatrix}
        4N + 1 & 1 & 1 & 1 \\
        (2N-1)(N+1) & 5N - 1 & - N - 1 & - N - 1 \\ 
        2N(2N-1) & - 2N & 4N & -2N \\
        N(2N-1) & -N & -N & 5N
    \end{pmatrix}.
\end{align}
For any integer $N$, we find $\mathrm{dim}\, \mathrm{ker}\, F = 1$, implying that only one of the four interaction terms in Eq.~\eqref{eq:basiseucl} is linearly independent. In particular, the kernel of $F$ is spanned by the vector $K = (-1,N+1,2N,N)^T$, and hence, for $N_f = 1$, the interacting Lagrangian in Eq.~\eqref{eq:basiseucl} can be rewritten in terms of a single interaction term
\begin{align}\label{eq:Nf1ints}
    \mathcal{L}_\mathrm{int} = - \frac{\bar{\tilde{g}}}{8} \big[&-(\bar{\chi}(\mathbb{1}_{2N} \otimes \mathbbm{1}_2)\chi)^2 + (N+1)(\bar{\chi} (\mathbb{S}_a \otimes \mathbbm{1}_2) \chi)^2 \nonumber \\
    &+2N(\bar{\chi} (\mathbb{A}_b \otimes \gamma_{\mu}) \chi)^2+N(\bar{\chi} (\mathbb{A}_b \otimes \gamma_5) \chi)^2 \big]
\end{align}
where 
\begin{equation}\label{eq:gtilde}
    \bar{\tilde{g}} = -\bar{g}_1 + (N+1) \bar{g}_2 + 2N\bar{g}_3 + N\bar{g}_4.
\end{equation}

\subsection{\texorpdfstring{$\beta$}{beta} functions}

\begin{figure*}[t!]
    \centering
    \includegraphics{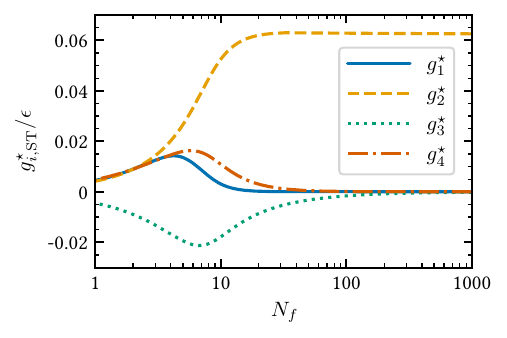}
    \includegraphics{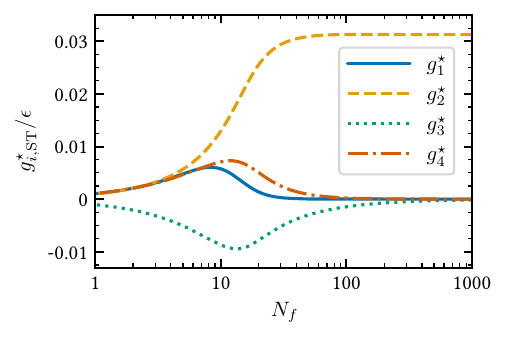}
    \caption{Fixed-point coordinates of the $\mathrm{SO}(2N)$-symmetric-tensor fixed point as given in Eq.~\eqref{eq:FPsymCoord} for $N=4$ (left) and $N=8$ (right). The fixed point separates a $\mathrm{SO}(2N)$--symmetric phase and an ordered phase that breaks down ${\mathrm{SO}(2N) \to \mathrm{SO}(N) \times \mathrm{SO}(N)}$.
    While $N=8$ correspond to charge-neutral twisted bilayer graphene~\cite{bistritzer2011moire}, $N=4$ corresponds to a single sheet of graphene.}
    \label{fig:stfpcoords}
\end{figure*}

We use the \texttt{MaRTIn}~\cite{Brod:2024zaz} toolkit to compute the $\overline{\mathrm{MS}}$ $\beta$ functions and fermion anomalous dimension of the theory defined by the Euclidean action
\begin{align}\label{eq:action}
    S = \int d^dx \left( \frac{1}{2} \bar{\chi} (\mathbb{1}_{2N} \otimes \gamma_\mu) \partial_\mu \chi + \mathcal{L}_\mathrm{int} \right)
\end{align}
in $d=2+\epsilon$ spacetime dimensions up to leading order, i.e., one-loop in the $\beta$ functions and two-loop in the fermion anomalous dimension.
We refer to App.~\ref{app:martin} for the computational details.

For convenience, we use the rescaled couplings, ${g_n := \bar{g}_n/(4\pi N_f)}$, and find the resulting one-loop $\beta$ functions $\beta_n(\epsilon) := \tfrac{d g_n}{d \log \mu}$ to be
\begin{align}
    \beta_1 =& \epsilon g_1 + \frac{1}{N_f} \bigg[-4(N N_f - 1)g_1^2 + 4 (N + 1)(2N - 1) g_1 g_2 \nonumber \\ 
    &\hspace{1cm}- 4N(2N - 1) g_1 g_4 + 8N(2N - 1) g_1 g_3 \notag \\
    &\hspace{1cm}- 8N(2N - 1) g_3 g_4 \bigg], \label{eq:beta1} \\
    \beta_2 =& \epsilon g_2 + \frac{1}{N_f} \bigg[-(4N (N_f - 1) + 4) g_2^2 + 4 g_1 g_2 \nonumber \\
    &\hspace{1cm}+ 8N(N-1) g_2 g_3 - 4N g_2 g_4 + 8N(N-1) g_3 g_4\bigg], \\
    \beta_3 =& \epsilon g_3 + \frac{1}{N_f} \bigg[8 N(N - 1) g_3^2 + 2N(N-1) g_4^2 + 2N(N + 1) g_2^2 \nonumber \\
    &\hspace{1cm} + 4(N+1)(N-1) g_2 g_4 + 4 g_1 g_4  \bigg], \\
    \beta_4 =& \epsilon g_4 + \frac{1}{N_f} \bigg[ 4N(N_f - 1) g_4^2 + 4(N+1) g_2 g_4 + 8 g_1 g_3 \nonumber \\
    &\hspace{1cm} + 8(N+1)(N-1) g_2 g_3  - 4 g_1 g_4 + 8N^2 g_3 g_4 \bigg] \label{eq:beta4} \,.
\end{align}
The sign of the $\beta$ functions is chosen such that if $\beta_n > 0$, the coupling $g_n$ decreases toward the IR. 

For $N_f = 1$, the $\beta$ functions above can be combined to a single $\beta$ function for $\tilde{g} = \bar{\tilde{g}}/(4 \pi N_f)$ with $\bar{\tilde{g}}$ defined as in Eq.~\eqref{eq:gtilde}, which coincides with the $\beta$ function of the canonical Gross--Neveu model~\cite{Gross:1974}
\begin{align}
    \beta_{\tilde{g}} = \epsilon \tilde{g} + 4(N-1)\tilde{g}^2.
\end{align}
Hence, for $N_f = 1$, where the four couplings become redundant, any nontrivial fixed point either corresponds to the Gaussian fixed point or the Gross--Neveu fixed point, given by
\begin{align}
    \tilde{g}^\star = -\frac{1}{4(N-1)} \epsilon\,,
\end{align}

\subsection{Fixed-point structure}

\begin{figure*}[t!]
    \centering
    \includegraphics{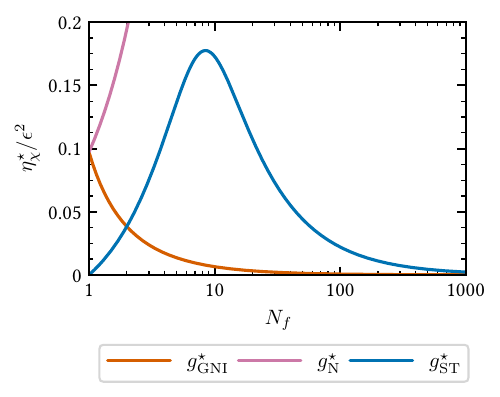}
    \includegraphics{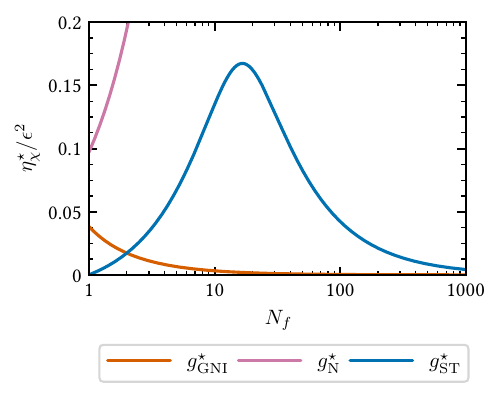}
    \caption{Anomalous dimension $\eta_\chi$ of the Majorana fermion for the $\mathrm{SO}(2N)$--symmetric theory in Eq.~\eqref{eq:action} at $N=4$ (left) and $N=8$ (right).
    The anomalous dimensions are evaluated at the Gross--Neveu--Ising (GNI)~\eqref{eq:IsingFP}, symmetric-tensor (ST)~\eqref{eq:FPsymCoord}, and adjoint-nematic fixed point (N)~\eqref{eq:nematicFP}.}
    \label{fig:anomdimSTEta}
\end{figure*}

In order to find critical points of the theory defined by Eq.~\eqref{eq:action}, we have to identify IR stable fixed points of the $\beta$ functions calculated above. Such an analysis majorly simplifies if one-loop closed subspaces can be identified, i.e., a subspace of the coupling space where additional couplings do not get generated under renormalization-group transformations. We find two of such one-loop closed subspaces:

\paragraph{Gross--Neveu--Ising subspace:} Since no term $\sim g_1^2$ appears in the flow equations of $g_n$ with $n = 2,3,4$, the subspace defined by $g_2 = g_3 = g_4 = 0$ is one-loop closed, see also, e.g., Refs.~\cite{Gehring:2015,Ladovrechis:2023,Hawashin:2025}. This ``Gross--Neveu--Ising subspace'' includes an interacting fixed-point, given by
\begin{align}\label{eq:IsingFP}
    g^\star_\mathrm{GNI} = \left[\frac{N_f}{4(N N_f - 1)},0,0,0\right] \epsilon,
\end{align}
which is IR stable for all $N_f > 1$. 
We remark that in the Gross--Neveu--Ising subspace for $N_f=1$, Eq.~\eqref{eq:gtilde} yields ${\tilde{g} = -g_1}$, and hence $g^\star_\mathrm{GNI} = - \tilde{g}^\star$.
Starting the flow with a small $g_1 > g_{\mathrm{GNI},1}^\star$ leads to a runaway flow to positive infinity, which corresponds to quantum anomalous Hall GNI criticality with order parameter $\langle \bar{\chi} \left(\mathbb{1}_{2N} \otimes \mathbb{1} \right) \chi \rangle > 0$, which spontaneously breaks parity as well as $\mathbb{Z}_2$ chiral symmetry.

\paragraph{Adjoint-nematic subspace:} Another fixed point appears in the one-loop closed subspace where only $g_3 \neq 0$, given by
\begin{align}\label{eq:nematicFP}
    g^\star_{\mathrm{N}} = \left[0,0,-\frac{N_f}{8 N(N - 1)},0\right] \epsilon.
\end{align}
It corresponds to adjoint-nematic quantum criticality with order parameter $\langle \bar{\chi} (\mathbb{A}_a \otimes \gamma_\mu)\chi\rangle > 0$. It is the lower-dimensional analog of the nematic critical point identified at and beyond mean-field in $2+1$ dimensions~\cite{Herbut:2023une,Han:2024ird}. Note that $g^\star_{\mathrm{N}} \sim N_f$, and hence the fixed point moves toward infinity in the limit $N_f \to \infty$. The same is true for all $N_f$ in the limit $N \to 1^+$, for which the corresponding interaction coincides with the Thirring coupling~\cite{Bondi:1989gba}.
The absence of a fixed-point for $N > 1$ that lies purely on the axis defined by the Thirring-like coupling $g_3$ in the large-$N_f$ limit is related to the vanishing $\beta$ function of the Thirring coupling at all loop-orders in $d=1+1$ dimensions for irreducible representations of the Lorentz group~\cite{BONDI1989345}.
The appearance of a $g_3^2/N_f$ term at finite $N_f$ and $N > 1$ is related to the multiplicity of the degrees of freedom, as also observed in related systems~\cite{Gehring:2015,Ladovrechis:2023}.
For $N_f = 1$, $g^\star_{\mathrm{N}}$ becomes equal to the GNI fixed point, 
i.e.,
\begin{equation}
    \tilde{g}(g^\star_\mathrm{N}) = -\frac{1}{4(N-1)} \epsilon \,,
\end{equation}
as directly follows from Eq.~\eqref{eq:gtilde}.
Notably, this differs from the result obtained with Wilsonian RG in $d=2+1$~\cite{Han:2024ird}, where the adjoint-nematic fixed point coincides with the Gaussian fixed point for $N_f = 1$.

We are also interested in fixed points that correspond to symmetric-tensor criticality, described by the order parameter $\langle \bar{\chi} (\mathbb{S}_a \otimes \mathbbm{1}_2) \chi \rangle$. Note, however, that a finite $g_2$ always induces a flow of $g_3$, and together they lead to a nontrivial flow of $g_1$ and $g_4$. Hence, there is no subspace that includes such a fixed point, and the full coupling space has to be analyzed. The identification of a suitable fixed point simplifies in the large-$N_f$ limit of the one-loop $\beta$ functions, where the individual $\beta$ functions decouple and reduce to 
\begin{align}
    \beta_1 &= \epsilon g_1 - 4N g_1^2 + \mathcal{O}(1/N_f), \\
    \beta_2 &= \epsilon g_2 - 4Ng_2^2 + \mathcal{O}(1/N_f), \\
    \beta_3 &= \epsilon g_3 + \mathcal{O}(1/N_f), \\
    \beta_4 &= \epsilon g_4 + 4Ng_4^2 + \mathcal{O}(1/N_f).
\end{align}
Note again the absence of a $g_3^2$ term in the large-$N_f$ limit, since $g^\star_\mathrm{N} \to \infty$. We can now clearly identify the IR stable fixed point 
\begin{align}
    g_\mathrm{ST}^\star = \left[\mathcal{O}(1/N_f^4), \frac{1}{4N} + \mathcal{O}(1/N_f),\mathcal{O}(1/N_f), \mathcal{O}(1/N_f^2)\right]\epsilon\,,
\end{align}
corresponding to symmetric-tensor criticality with the symmetric tensor order parameter $\langle \bar{\chi} (\mathbb{S}_a \otimes \mathbbm{1}_2) \chi \rangle > 0$, which signals spontaneous symmetry breaking of $\mathrm{SO}(2N) \to \mathrm{SO}(N) \times \mathrm{SO}(N)$. 
Including finite $N_f$ corrections, the fixed-point coordinates will be a complicated function of $N_f$, and we set
\begin{equation}\label{eq:FPsymCoord}
    g_\mathrm{ST}^\star=[h_1(N_f),h_2(N_f),h_3(N_f),h_4(N_f)]\epsilon \,.
\end{equation}
Up to $\mathcal{O}(1/N_f^3)$, the functions $h_i$ read
\begin{align}
    h_1(N_f) &=  \mathcal{O}(1/N_f^4),\\
    h_2(N_f) &= \frac{1}{4N} + \frac{N-1}{4N^2} \frac{1}{N_f} - \frac{(N^3 + N^2 - N + 1)(N-1)}{4N^3} \frac{1}{N_f^2} \notag\\ &\hspace{0.5cm}+ \mathcal{O}(1/N_f^3),\\
    h_3(N_f) &= -\frac{N + 1}{8N} \frac{1}{N_f} - \frac{(N+1)(N-1)}{4N^2} \frac{1}{N_f^2} \notag\\ &\hspace{0.5cm}+ \mathcal{O}(1/N_f^3), \\
    h_4(N_f) &= -\frac{(N-1)(N+1)^2}{4N^2} \frac{1}{N_f^2} + \mathcal{O}(1/N_f^3).
\end{align}
We show the dependence of the couplings on $N_f$ for the case relevant to graphene ($N = 4$) as well as twisted bilayer graphene ($N=8$) in Fig.~\ref{fig:stfpcoords}. For $N_f = 1$, we find that $\tilde{g}(g^\star_\mathrm{ST}) = 0$, i.e., it becomes equivalent to the Gaussian fixed point. This agrees with the findings from Wilsonian RG directly in $d=2+1$~\cite{Han:2024ird}.

\section{Critical exponents} \label{sec:critExp}

\begin{figure*}[t!]
    \centering
    \includegraphics{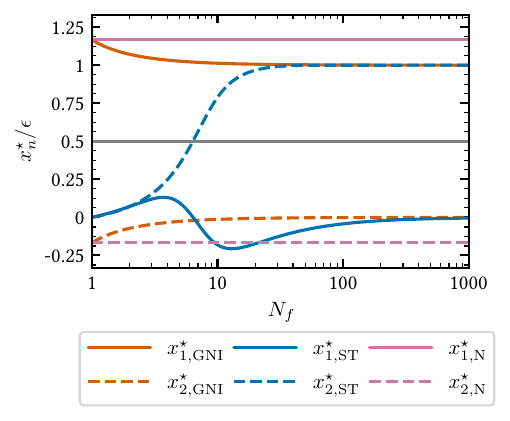}
    \includegraphics{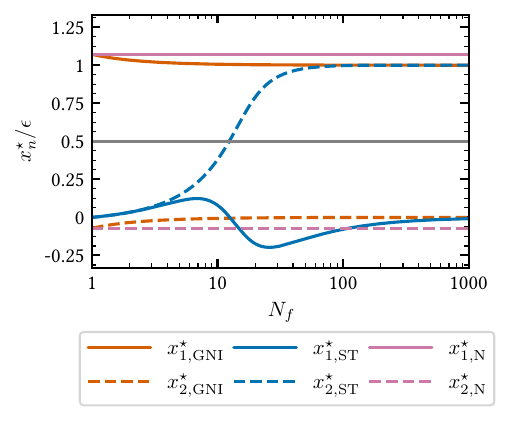}
    \caption{Anomalous dimension $\eta_{\phi_n}$ of the Gross--Neveu--Ising, $\phi_1 = \langle \bar{\chi} (\mathbbm{1}_{2N} \otimes \mathbbm{1}_2) \chi \rangle$, symmetric tensor, $\phi_2 = \langle \bar{\chi} (\mathbb{S}_a \otimes \mathbbm{1}_2) \chi \rangle$, and adjoint nematic, $\phi_3 = \langle \bar{\chi} (\mathbb{A}_b \otimes \gamma_\mu \chi \rangle$, order parameters for the $\mathrm{SO}(2N)$--symmetric theory in Eq.~\eqref{eq:action}.
    We show the values of $x_n = \tfrac{1}{2} \left( 2 + \epsilon - \eta_{\phi_n} \right)$ for the $\mathrm{SO}(8)$-- (left) and $\mathrm{SO}(16)$--symmetric theory (right). 
    At the one-loop order, $\phi_3$ receives no quantum corrections, i.e., $\eta_{\phi_3}=2+\epsilon$, so the corresponding value of $x_3 = 0$ is omitted from the figures for clarity.
    At a critical point, the susceptibility of the corresponding order parameter $\phi_n$ diverges, requiring $x_n > \epsilon/2$, which is marked as a solid gray line.
    All anomalous dimensions are evaluated at the Gross--Neveu--Ising (GNI)~\eqref{eq:IsingFP}, symmetric-tensor (ST)~\eqref{eq:FPsymCoord}, and adjoint-nematic fixed point (N)~\eqref{eq:nematicFP}.
    While, at the symmetric-tensor fixed point, $x_1$ and $x_3$ never imply critical behavior, $x_2$ implies criticality only for $N_f > N_{f,c}^{\mathrm{ST}}(N=4) \approx 6.3$, or $N_f > N_{f,c}^{\mathrm{ST}}(N=8) \approx 12.5$ respectively. Probing the condition $x_2 > \epsilon/2$ for various integer values of $N$, we find $N_{f,c}^{\mathrm{ST}}(N) \approx 0.56 + 1.48 N +\mathcal{O}(\epsilon)$.}
    \label{fig:anomdimSTOP}
\end{figure*}

\subsection{Correlation-length exponent}

The correlation-length exponent $\nu$ is given by the inverse of the largest positive eigenvalue of the stability matrix ${S_{ij} = - \partial \beta_i / \partial g_j}$. At one-loop order, there is a unique positive eigenvalue given by $\epsilon$~\cite{Gehring:2015}, and hence
\begin{equation}
    1/\nu = \epsilon + \mathcal{O}(\epsilon^2).
\end{equation}

\subsection{Fermion anomalous dimension}

The fermion anomalous dimension vanishes at one-loop order in any four-fermion theory close to two spacetime dimensions, i.e., $\eta_\chi(g^{\star}) = \mathcal{O}(\epsilon^2)$. This, on the other hand, implies that the one-loop $\beta$ functions for the couplings together with the two-loop result for the anomalous dimension are sufficient to compute the fermion anomalous dimension to $\mathcal{O}(\epsilon^2)$. 
We find the fermion anomalous dimension $\eta_\chi := \frac{d \log Z_\chi}{d \log \mu}$ to be
\begin{align}\label{eq:fermADM}
    \eta_\chi = &\frac{1}{N_f^2} \Bigg[ -8N(2N - 1)\, g_1 g_3 + 8 N^2(2N - 1)N\, g_3 g_4 \notag \\
    &- 4N(2N - 1)\, g_1 g_4 - 4(N+1)(2N-1)\, g_1 g_2  \notag \\
    &+ 8N(N+1)(2N-1)\, g_2 g_3 + 4N(N+1)(2N-1) \, g_2 g_4 \notag \\
    &+ 8 N_f N^2(2N - 1) \, g_3^2 + 2 N^2(2N - 1)(2N_f - 1) \, g_4^2 \notag \\
    &+ 2 (N+1)(2N-1)\,(N(2N_f - 1) + 1) \, g_2^2 \notag \\
    &+ 2(2N N_f - 1)\, g_1^2 \Bigg]\,,
\end{align}
where $Z_\chi$ is the field renormalization constant of the Majorana fermion, cf. App.~\ref{app:martin} for details.

Evaluating the expression above at the GNI fixed point~\eqref{eq:IsingFP} yields
\begin{align}
    \eta_\chi^\mathrm{GNI} = \frac{2N N_f - 1}{8(N N_f - 1)^2} \epsilon^2 + \mathcal{O}(\epsilon^3).
\end{align}
For $N_f = 1$, this coincides with the standard result for the Gross--Neveu-Ising universality class~\cite{Bondi:1989gba,Gehring:2015,Ladovrechis:2023}, and further agrees with the result from the large-$N$ expansion~\cite{Gracey:2025aoj} up to $\mathcal{O}(1/N^4)$ when expanding it in powers of $\epsilon = d - 2$.

The fermion anomalous dimension at the symmetric-tensor fixed point~\eqref{eq:FPsymCoord} up to $\mathcal{O}(1/N_f^3,\epsilon^2)$ is given by
\begin{align}
    \eta_\chi^\mathrm{ST} = &\left(\frac{4N(2N-1)(N+1)}{N_f} + \frac{6(N-1)(N+1)(2N-1)}{N_f^2}\right)\epsilon \notag \\
    &+\mathcal{O}(1/N_f^3,\epsilon^2).
\end{align}
In Fig.~\ref{fig:anomdimSTEta}, we show the full dependence of $\eta_\chi$ on $N_f$ evaluated at the symmetric-tensor fixed point. As can be seen, it vanishes for $N_f = 1$, confirming again its equivalence with the Gaussian fixed point. 

For the adjoint-nematic fixed point~\eqref{eq:nematicFP}, we find
\begin{align}
    \eta_\chi^\mathrm{N} = \frac{N_f(2N - 1)}{8(N-1)^2} \epsilon^2 + \mathcal{O}(\epsilon^3).
\end{align}
Precisely for $N_f = 1$ it coincides with $\eta_\psi^\mathrm{GNI}$, as noted earlier, due to the equivalence of the adjoint-nematic fixed point and the GNI fixed point.

\subsection{Order parameter anomalous dimension}

In order to calculate the anomalous dimension of the order parameters corresponding to Gross--Neveu, adjoint-nematic and symmetric-tensor criticality, we do a susceptibility analysis following the prescription of Ref.~\cite{Janssen:2016}. To that end, we perturb the action in Eq.~\eqref{eq:action} by the symmetry-breaking term
\begin{align}\label{eq:OPanomdim}
    S \to S + y_n \int d^dx \bar{\chi} \mathcal{M}_n \chi
\end{align}
with $n = 1,2,3$ and $\mathcal{M}_1 = \mathbb{1}_{2N} \otimes \mathbb{1}_2$, $\mathcal{M}_2 = \mathbb{S}_a \otimes \mathbb{1}_2$, and $\mathcal{M}_3 = \mathbb{A}_b \otimes \gamma_\mu$ for fixed $a$, $b$ and $\mu$. 
The corresponding $\beta$ function will be given by
\begin{equation}
    \beta_{y_n} = \frac{d y_n}{d \log \mu} = - (1+x_n) y_n,
\end{equation}
where $x_n$ depends on all other couplings of the theory space~\eqref{eq:basiseucl}. 

The anomalous dimension of the corresponding order parameters $\phi_n = \langle \bar{\chi} \mathcal{M}_n \chi \rangle$ then follows from the scaling of the susceptibility together with the hyperscaling relation~\cite{Han:2024ird,Janssen:2016,Ladovrechis:2023} $\eta_\phi = 2 - \gamma/\nu$ as 
\begin{equation}\label{eq:etaphiN}
    \eta_{\phi_n} = 2 + \epsilon - 2x_n.
\end{equation}
We find at the one-loop order
\begin{align}
    \eta_{\phi_1} &= 2 + \epsilon - \left( \frac{4}{N_f} - 8N \right)\, g_1 + \frac{4(N + 1) (2N - 1)}{N_f}\, g_2 \notag\\ 
                    &+ \frac{8N(2N-1)}{N_f}\, g_3  + \frac{4N(2N-1)}{N_f}\, g_4 \,, \label{eq:etaphi1}\\
    \eta_{\phi_2} &= 2 + \epsilon + \frac{4}{N_f}\, g_1 - \frac{4(2N N_f - N + 1)}{N_f}\, g_2 \notag\\
                    & - \frac{8N}{N_f}\, g_3 + \frac{4N}{N_f}\, g_4 \,, \label{eq:etaphi2} \\
\eta_{\phi_3} &= 2 + \epsilon \,. \label{eq:etaphi3} 
\end{align}
Note that at a true critical point, the corresponding susceptibility is diverging, implying $\gamma_n/\nu_n > 0$. For this to hold in $d = 2+\epsilon$, we have to have $x_n > \frac{\epsilon}{2}$, i.e., $x_n > 1/2$ for $\epsilon = 1$, and equivalently $\eta_\phi < 2$. 

At the Gross--Neveu-Ising fixed point~\eqref{eq:IsingFP}, the anomalous dimension of the Ising order parameter is given by
\begin{align}
    \eta_{\phi_1}^\mathrm{GNI} = 2 - \frac{N N_f}{NN_f - 1} \epsilon\,,
\end{align}
which is in agreement with the previous literature~\cite{Ladovrechis:2023,Gracey:2025aoj}. 

The anomalous dimension of the symmetric-tensor order parameter reads up to order $\mathcal{O}(1/N_f^3,\epsilon^2)$
\begin{align}
    \eta_{\phi_2}^{\mathrm{ST}} = &2 + \bigg[1-8N - 4 \frac{N-1}{N_f} \notag\\
    &+ \frac{4(1+N(2N-1))(N^2+N-1)}{N N_f^2}\bigg]\epsilon \notag\\
    &+ \mathcal{O}(1/N_f^3,\epsilon^2)
\end{align}

For the adjoint-nematic fixed point~\eqref{eq:nematicFP}, we do not find any correction to the leading order at one-loop, i.e.,
\begin{align}
    \eta_{\phi_3}^\mathrm{N} = 2 + \epsilon \,.
\end{align}
In Fig.~\ref{fig:anomdimSTOP}, we show the order parameter anomalous dimensions of the GNI $\phi_1$ and symmetric-tensor $\phi_2$ order parameter as a function of $N_f \geq 1$ and for $N=4,8$, evaluated at the GNI~\eqref{eq:IsingFP}, symmetric-tensor~\eqref{eq:FPsymCoord}, and adjoint-nematic fixed 
point~\eqref{eq:nematicFP}. While the GNI fixed point remains critical for all $N_f$, the symmetric-tensor fixed point implies a diverging susceptibility in the symmetric-tensor channel only for $N_f > N_{f,c}^{\mathrm{ST}}(N=4) \approx 6.3$, or $N_f > N_{f,c}^{\mathrm{ST}}(N=8) \approx 12.5$, respectively. Directly in $d=2+1$, Ref.~\cite{Han:2024ird} finds $N_{f,c}^{\mathrm{ST}}(N=4) \approx 1.88$, which is much lower than our estimate. Note, however, that the Wilsonian RG analysis of Ref.~\cite{Han:2024ird} is well controlled only in the large-$N_f$ limit, and that a naive extrapolation of our leading-order result is not sufficient to expect agreement. Repeating the analysis for the symmetric-tensor fixed point for various values $N \in \{2, 4, 8, 10, 50, 100, 500\}$ allows for a linear fit of $N_{f,c}^{\mathrm{ST}}(N)$, yielding
\begin{equation}\label{eq:NfcFit}
    N_{f,c}^{\mathrm{ST}}(N) \approx 0.56 + 1.48 N +\mathcal{O}(\epsilon)\,.
\end{equation}
The continuity of the symmetric-tensor susceptibility along the symmetric-tensor transition for $N_f < N_{f,c}^{\mathrm{ST}}$ suggest that the transition is first-order, in agreement with Refs.~\cite{Han:2024ird,Han:2024swe,Han:2025kjt}.

\subsection{Padé interpolation of \texorpdfstring{$N_{f,c}$}{Nfc} with \texorpdfstring{$4-\varepsilon$ }{4-epsilon} results}

An improved estimate for $N_{f,c}$ in $d=2+1$ dimensions can be obtained by interpolating our one-loop result for $N_{f,c}$ from the lower critical dimension with the three-loop estimate in the complementary $\mathrm{SO}(2N)$--symmetric Gross--Neveu--Yukawa theory studied in Ref.~\cite{Han:2025kjt} from the upper critical dimension $d=4-\varepsilon$ by using Padé approximants. 
Note that this is a rather heuristic approach, since the mechanisms leading to a finite $N_{f,c}$ close to the two critical dimensions are different. Such a two-sided Padé interpolation of critical numbers of field components has been already employed in the context of fixed-point collision in the Abelian Higgs model~\cite{Ihrig:2018}, but also for the interpolation of various critical exponents in Gross--Neveu models~\cite{Ihrig:2018,Ladovrechis:2023,Hawashin:2025,Gracey:2025aoj}. Padé approximants are a class of functions defined by
\begin{align}
    [m/n](d) = \frac{\sum_{i=0}^m a_i d^i}{1 + \sum_{i=1}^n b_i d^i}\,,
\end{align}
which are parametrized by $m+n+1$ coefficients $a_i$ and $b_i$. Here, we will choose the coefficients $a_i$ and $b_i$ such that $[m/n](2+\epsilon)$ matches our result for $N_{f,c}^{\mathrm{ST}}$ in Eq.~\eqref{eq:NfcFit} up to zeroth order in $\epsilon$, and $[m/n](4-\varepsilon)$ matches the results from $N_{f,c}$ up second order in $\varepsilon$.

For $N=4$ and $N=8$,\footnote{We thank SangEun Han and Igor Herbut for providing their estimate of $N_{f,c}$ for $N=8$ beyond the large-$N$ formula in Ref.~\cite{Han:2025kjt} in private communication.} Ref.~\cite{Han:2025kjt} finds up to three-loop order\footnote{Note that the parameter $N$ appearing in Ref.~\cite{Han:2025kjt} is related to the parameter $N$ of this work via $N^{\mathrm{Ref.}~[26]} = 2N^{\text{this work}}$.}
\begin{align}
    N_{f,c}^{\mathrm{ST}, \mathrm{Ref.}~[26]}(N=4) &\approx 10.798 - 3.619\varepsilon + 4.640\varepsilon^2 + \mathcal{O}(\varepsilon^3) \,, \\
    N_{f,c}^{\mathrm{ST}, \mathrm{Ref.}~[26]}(N=8) &\approx 16.444 - 3.737\varepsilon + 10.202\varepsilon^2 + \mathcal{O}(\varepsilon^3) \,,
\end{align}
This yields four constraints, implying that we can determine all Padé approximants with $m+n+1 = 4$. 

We find $N_{f,c}(N=4) \approx 8.9$ and $N_{f,c}(N=8) \approx 16.1$ by averaging over all two-sided Padé approximants, which compares well with the Borel--Padé-resummed results of Ref.~\cite{Han:2025kjt}.

\section{Conclusion} \label{sec:conclusion}

In this work, we have studied spontaneous symmetry breaking of $\mathrm{SO}(2N)$ in the paradigmatic Gross--Neveu theory. First, we have shown that if certain interactions are tuned to zero, the corresponding restricted Gross--Neveu theory space has a symmetry enhancement from $\mathrm{SU}(N) \times \mathrm{U}(1)$ to $\mathrm{SO}(2N)$. To that end, we considered a particular class of irreducible representations of the Clifford algebra, which allowed us to make the $\mathrm{SO}(2N)$ symmetry manifest if the theory is rewritten in terms of $4N$-component Majorana fermions. We have then extended the theory by an additional flavor degree of freedom, and constructed a Fierz-complete renormalizable Lagrangian that includes four independent couplings. We have further shown by employing the Fierz rearrangement theorem that in the case of trivial flavor structure, the four interactions become redundant and ultimately reduce to one linear-independent four-fermion interaction that is equivalent to the canonical Gross--Neveu interaction. Furthermore, we computed the one–loop $\beta$ functions together with the one-loop order parameter and two–loop fermion anomalous dimensions. We found three fixed points that are the lower-dimensional analogs to the fixed points found in the previous $d=2+1$ Wilsonian RG analysis~\cite{Han:2024ird}: (i)~QAH Gross--Neveu--Ising, (ii)~symmetric–tensor, and (iii)~adjoint–nematic fixed points. 
While (i) and (iii) are located within one-loop closed subspaces, (ii) lives in the full coupling space but is tractable in a controlled large–$N_f$ limit.

We have shown that any of these fixed points survive all the way down to $N_f = 1$. In particular, for $N_f = 1$, the fixed point (ii) becomes identical to the Gaussian fixed point, while the fixed-point (iii) coincides with (i). For the fixed point (iii), we do not find a diverging susceptibility for all $N$ and $N_f \geq 1$ at leading order, while the susceptibility corresponding to symmetric-tensor criticality diverges only for theories where $N_f > N_{f,c}^{\mathrm{ST}} \approx 0.56 + 1.48 N$. Using a two-sided Padé interpolation that takes both estimates for $N_{f,c}^{\mathrm{ST}}$ from the upper and lower critical dimension into account, we find $N_{f,c}^{\mathrm{ST}}(N=4) \approx 8.9$ and $N_{f,c}(N=8) \approx 16.1$. 

After the publication of our work, Ref.~\cite{Rein:2025cje} reported quantum Monte Carlo evidence that the putative $\mathrm{SO}(8)$ symmetric-tensor transition of the canonical Gross--Neveu model is likely already weakly first order. Moreover, they observe that the discontinuity increases with $N$, suggesting an increasingly robust first-order transition. Our results are in agreement with the findings of Ref.~\cite{Han:2024ird,PhysRevB.110.125131,Rein:2025cje}, and supports the interpretation that the canonical Gross--Neveu model features a second-order transition only in the Gross--Neveu--Ising channel. 

For future work, it would be interesting to investigate the effect of higher-loop corrections to our finding. It would be also interesting to better understand the absence of a diverging susceptibility at a fixed point, and connect it to the picture from the equivalent Gross--Neveu--Yukawa theory, where the absence of a continuous transition can be understood in a more conventional way as a fixed point collision. It would also be interesting to corroborate our findings by a non-perturbative renormalization group analysis.

\begin{acknowledgments}
We thank SangEun Han, Igor Herbut, Michael Scherer, Emmanuel Stamou and Tom Steudtner for valuable discussions, suggestions, and comments on the manuscript.
We are supported by the Mercator Research Center Ruhr under Project No.~Ko-2022-0012. M.U.~is supported by the doctoral scholarship program
of the \textit{Studienstiftung des deutschen Volkes}.
\end{acknowledgments}

\appendix

\section{Computational Details in Minkowskian spacetime}\label{app:martin}

In this Appendix, we provide further details on the computation of the renormalization group equations given in the main text, namely the one-loop $\beta$ functions~\eqref{eq:beta1}--\eqref{eq:beta4}, the one-loop order parameter anomalous dimensions~\eqref{eq:etaphi1}--\eqref{eq:etaphi3}, and the two-loop fermion anomalous dimension~\eqref{eq:fermADM}.

We employ dimensional regularization~\cite{Bollini:1972ui}, the modified minimal subtraction ($\overline{\mathrm{MS}}$) scheme, and analytically continue the Euclidean-signature Lagrangian~\eqref{eq:action} to $d=2+\epsilon$ Minkowskian spacetime dimensions.
We start with a bare Lagrangian comprised of $N_f$ copies of $\mathrm{SO}(2N)$--vector Majorana fermions $\chi_{i, k}$ with $i,j=1,\dots,2N$ and $k=1,\dots,N_f$, i.e.,
\begin{align}\label{eq:bareLagrMinkowski}
    \mathcal{L}^{(0)} &= \frac{1}{2} \bar{\chi}_{i, k}^{(0)} i \partial_\mu \gamma^\mu \chi^{(0)}_{i, k} + \sum_{n=0}^2 \Bigg\{ 
    \frac{ g^{(0)}_{A,n}}{8} \left( \bar{\chi}^{(0)}_{i, k}  \Gamma_{(n)}^{\mu_1 \ldots \mu_n} \chi^{(0)}_{i, k} \right)^2 \notag\\
    &+ \frac{ g^{(0)}_{B,n}}{8} \left( \bar{\chi}^{(0)}_{i, k} \Gamma_{(n)}^{\mu_1 \ldots \mu_n} \chi^{(0)}_{j, k} \right) \left( \bar{\chi}^{(0)}_{j, k} \Gamma_{(n)\,\mu_1 \ldots \mu_n}  \chi^{(0)}_{i, k} \right)  \notag\\
     &+ \frac{ g^{(0)}_{C,n}}{8} \left( \bar{\chi}^{(0)}_{i, k} \Gamma_{(n)}^{\mu_1 \ldots \mu_n} \chi^{(0)}_{j, k} \right) \left( \bar{\chi}^{(0)}_{i, k} \Gamma_{(n)\,\mu_1 \ldots \mu_n} \chi^{(0)}_{j, k} \right) 
    \Bigg\} \,,
\end{align}
where the superscript ``$(0)$'' denotes bare quantities, and the $\Gamma_{(n)}$ are the Dirac structures 
\begin{equation}\label{eq:diracStructures}
    \Gamma_{(0)} = \mathbbm{1}_2 \,,\quad \Gamma_{(1)}^{\mu} = \gamma^\mu \,,\quad \Gamma_{(2)}^{\mu \nu} = \frac{1}{2} ( \gamma^\mu \gamma^\nu - \gamma^\nu \gamma^\mu) \,.
\end{equation}
The Lagrangian~\eqref{eq:bareLagrMinkowski} can be mapped to the Euclidean-signature Lagrangian~\eqref{eq:action} by performing a Wick rotation, imposing a representation of the Hermitian gamma matrices where 
\begin{equation}\label{eq:gamReprEucl}
    \gamma_0^* = -\gamma_0\,,\quad \gamma_i^* = \gamma_i\,,\quad\mathrm{ for }\,\, i=1,5\,,
\end{equation}
and carrying out a change of basis from $\{g^{(0)}_{A,n},g^{(0)}_{B,n},g^{(0)}_{C,n}\}_{n=0,1,2}$ to $\{\bar{g}_1,\bar{g}_2,\bar{g}_3,\bar{g}_4\}$, see below.

The Dirac structures~\eqref{eq:diracStructures} constitute a basis of the $d=2$ Clifford algebra $\left\{\gamma^\mu, \gamma^\nu\right\}=2 \eta^{\mu \nu}$, and are thereby equivalent to the basis $\{\mathbb{1}_2,\gamma^{\mu}, \gamma_5 \}$ chosen in the main text, cf., Eq.~\eqref{eq:basis2D}.
Note that we opted for the basis element $\Gamma_{(2)}^{\mu \nu}$ in place of $\gamma_5$, which, in strictly $d=2$, are interrelated as $\Gamma_{(2)}^{\mu\nu} = \epsilon^{\mu\nu}\gamma_5$ with $\epsilon^{\mu\nu} \epsilon_{\mu\nu} = -2$. This choice circumvents an explicit treatment of $\gamma_5$, which, in dimensional regularization, necessitates the use of a suitable renormalization scheme like the 't Hooft--Veltman--Breitenlohner--Maison scheme~\cite{tHooft:1972tcz,Breitenlohner:1975hg,Ihrig:2019kfv,Steudtner:2025blh}.

For convenience, as compared to the Lagrangian~\eqref{eq:action}, we decomposed the antisymmetric matrices $\mathbb{A}^b$ and symmetric-traceless $\mathbb{S}^a$ into contractions of fundamental indices pertaining to the global $\mathrm{SO}(2N)$ group, yielding
\begin{align}
(\mathbb{S}^a)_{ij}(\mathbb{S}^a)_{kl}
&= \frac{c}{2}\big(\delta_{ik}\delta_{jl}+\delta_{il}\delta_{jk}\big)
  - \frac{c}{2N}\,\delta_{ij}\delta_{kl}\,, \\
(\mathbb{A}^b)_{ij}(\mathbb{A}^b)_{kl}
&= -\,\frac{c}{2}\big(\delta_{ik}\delta_{jl}-\delta_{il}\delta_{jk}\big)\,,
\end{align}
where $c$ denotes the normalization of the traces $\mathrm{Tr}(\mathbb{A}^a \mathbb{A}^{a^{\prime}}) = c \,\delta^{a a^{\prime}}$ and $\mathrm{Tr}(\mathbb{S}^b \mathbb{S}^{b^{\prime}}) = c \,\delta^{b b^{\prime}}$. 
Thus, the change of basis from the Lagrangian~\eqref{eq:bareLagrMinkowski} to the main text Lagrangian~\eqref{eq:action} is given by 
\begin{align}
\bar{g}_1 &= g_{A,0} + \frac{1}{2N}\,(g_{B,0}+g_{C,0}), &
\bar{g}_2 &= \frac{1}{c}\,(g_{B,0}+g_{C,0}), \notag \\
\bar{g}_3 &= \frac{1}{c}\,(g_{B,1}-g_{C,1}), & 
\bar{g}_4 &= \frac{1}{2c}\,(g_{C,2}-g_{B,2})\,, & 
\end{align}
where we suppressed the bare superscripts for brevity.
The trace normalization chosen in the main text corresponds to the value $c=2N$.

We remark that the Fierz-complete basis of symmetry-allowed interaction terms given by Eq.~\eqref{eq:basiseucl} does not depend on the choice of metric signature.
While for the Euclidean-signature Lagrangian~\eqref{eq:action}, we impose Eq.~\eqref{eq:gamReprEucl} for the three Hermitian $\gamma$-matrices, the Minkowskian-signature Lagrangian~\eqref{eq:bareLagrMinkowski} requires to take one of the $\gamma$-matrices to be anti-Hermitian. To retain manifest $\mathrm{SO}(2N)$ invariance and the same basis of four-fermion interactions, we have to choose, e.g., $\gamma_0^\dagger = - \gamma_0$ and $\gamma_i^\dagger = \gamma_i$ for $i=1,5$.

In dimensional regularization, a complication arises as the spacetime dimension $d$ becomes a continuous variable. Consequently, the Clifford algebra becomes infinite-dimensional, and identities such as $\Gamma_{(2)}^{\mu\nu} = \epsilon^{\mu\nu}\gamma_5$ are no longer valid. A common choice of basis in general $d$ dimensions is given by the fully antisymmetrized product of gamma matrices, i.e.,~\cite{Kennedy:1981kp,BONDI1989345,Vasiliev:1995qj,Gracey:2016mio}
\begin{align}
\Gamma_{(n)}^{\mu_1 \ldots \mu_n} =\gamma^{\left[\mu_1\right.} \ldots \gamma^{\left.\mu_n\right]} & =\frac{1}{n!} \sum_{\text {perm }}(-1)^p \gamma^{\mu_1} \ldots \gamma^{\mu_n} \,.
\end{align}
For integer dimensions $D$, the basis becomes finite-dimensional with $\Gamma_{(n)}^{\mu_1 \ldots \mu_n} = 0 \,,\; n > D$, owing to the antisymmetrization. At loop level, additional four-fermion operators, which include contractions of $\Gamma_{(n)}^{\mu_1 \ldots \mu_n}$ with $n \geq 3$, are dynamically generated. These so-called \textit{evanescent} operators are not part of the physical operator basis given in Eq.~\eqref{eq:bareLagrMinkowski} as they vanish when $\epsilon \to 0$. 

Thus, our renormalized Lagrangian in $d=2+\epsilon$ has to include evanescent four-fermion operators having $n \geq 3$ and reads
\begin{align}
    \Lagr &= \frac{1}{2} Z_\chi \bar{\chi}_{i,k} i \slashed \partial \chi_{i,k} \notag\\
    &+ \sum_{n=0}^\infty Z_\chi^2 \mu^{2\epsilon}  \Bigg\{ 
    \frac{ g_{A,n} + \delta g_{A,n}}{8} \left( \bar{\chi}_{i,k} \Gamma_{(n)}^{\mu_1 \ldots \mu_n} \chi_{i,k} \right)^2 \notag\\
    &+ \frac{ g_{B,n} + \delta g_{B,n}}{8} \left( \bar{\chi}_{i,k} \Gamma_{(n)}^{\mu_1 \ldots \mu_n} \chi_{j,k} \right) \left( \bar{\chi}^k_j \Gamma_{(n), \mu_1 \ldots \mu_n} \chi_{i,k} \right)  \notag\\
     &+ \frac{ g_{C,n} + \delta g_{C,n}}{8} \left( \bar{\chi}_{i,k} \Gamma_{(n)}^{\mu_1 \ldots \mu_n} \chi_{j,k} \right) \left( \bar{\chi}_{i,k} \Gamma_{(n), \mu_1 \ldots \mu_n} \chi_{j,k} \right) 
    \Bigg\},
\end{align}
whose renormalized fields and couplings are related to their bare counterparts~\eqref{eq:bareLagrMinkowski} via
\begin{align}
    \chi_{i, k}^{(0)} &= Z_{\chi}^{1/2} \chi_{i, k} \,,&
     g_{x, n}^{(0)} &= \mu^{-\epsilon} \left(g_{x, n} + \delta g_{x, n}\right),&&&&\label{eq:bareQuantities}
\end{align}
with $x \in \{A,B,C\}$.
Although, in principle, an infinite tower of evanescent operators has to be considered, a $L$-loop computation only induces a finite amount of operators satisfying $n \leq 3L + 2$~\cite{Bondi:1989gba}.

To obtain the order parameter anomalous dimensions~\eqref{eq:etaphiN}, we proceed analogously and renormalize the corresponding operators $y_n \bar{\chi} \mathcal{M}_n \chi$~\eqref{eq:OPanomdim}, with masslike coupling $[y_n] = 1$ via 
$y_n^{(0)} = \mu Z_{y_n} y_n$.

We carry out the renormalization in $d=2+\epsilon$ using the \texttt{MaRTIn} framework~\cite{Brod:2024zaz}. 
Internally, \texttt{MaRTIn} leverages the technique of infrared rearrangement~\cite{Chetyrkin_1998} to extract all UV divergent contributions of the relevant Green's functions from massive vacuum integrals. At one- and two-loop order, respectively, all appearing integrals can be reduced to a single master integral (given, e.g., in Ref.~\cite{Gracey:2016mio}). Throughout our loop computation, products of $n$ $\gamma$-matrices appear that necessitate decomposition and projection onto the basis of $\Gamma_{(n)}^{\mu_1 \ldots \mu_n}$. This is accomplished by the iterative identities~\cite{Vasiliev:1995qj,Gracey:2016mio}
\begin{align}
& \Gamma_{(n)}^{\mu_1 \ldots \mu_n} \gamma^\nu=\Gamma_{(n+1)}^{\mu_1 \ldots \mu_n \nu}+\sum_{r=1}^n(-1)^{n-r} \eta^{\mu_r \nu} \Gamma_{(n-1)}^{\mu_1 \ldots \mu_{r-1} \mu_{r+1} \ldots \mu_n}\,,\\
& \gamma^\nu \Gamma_{(n)}^{\mu_1 \ldots \mu_n}=\Gamma_{(n+1)}^{\nu \mu_1 \ldots \mu_n}+\sum_{r=1}^n(-1)^{r-1} \eta^{\mu_r \nu} \Gamma_{(n-1)}^{\mu_1 \ldots \mu_{r-1} \mu_{r+1} \ldots \mu_n} \,,
\end{align}
which for $n=1$ yield
\begin{equation}
\gamma^\mu \gamma^\nu= \Gamma_{(2)}^{\mu \nu} +\eta^{\mu \nu} \mathbbm{1} \,.
\end{equation}

At the one-loop order, no evanescent operators can mix into the $1/\epsilon$ poles of physical operators, i.e., the $\beta$ functions, because their insertions are of order $\mathcal{O}(\epsilon)$. Beginning at the two-loop order, their mixing becomes pertinent, and would require a suitable choice of renormalization scheme to extract the physical $\beta$ functions~\cite{BONDI1989345,Dugan:1990df,Gracey:2016mio}. The resulting $\beta$ functions of the physical couplings, $\beta_{g_{x,n}} := \tfrac{d g_{x,n}}{d \log \mu}$,  having $n=0,1,2$ with $x \in \{A,B,C\}$, and the anomalous dimension of the fermion field, $\eta_\chi := \tfrac{d \log Z_\chi}{d \log \mu}$, as well as the anomalous dimensions of the order parameters~\eqref{eq:etaphiN}
can be found as a Mathematica file in the
ancillary files of the arXiv submission of this work.
Our results have been corroborated against Refs.~\cite{Bondi:1989gba,BONDI1989345}.

\FloatBarrier
\bibliographystyle{JHEP}
\bibliography{references}

\end{document}